\documentstyle[prb,preprint,aps]{revtex}
\begin{document}
\draft
\title{Magnetic field of an in-plane vortex \\
inside and outside a layered superconducting film}
\author{Edson Sardella}
\address{Departamento de F\'{\i}sica,
Faculdade de Ci\^encias, Universidade Estadual Paulista \\
Caixa Postal 473, 17033-360, Bauru-SP, Brazil}
\date{\today}
\maketitle
\begin{abstract}
In the present work we study an anisotropic
layered superconducting film of finite thickness. The
film surfaces are
considered parallel to the $bc$ face of the crystal. The
vortex lines are oriented perpendicular to the film surfaces
and parallel to the superconducting planes.
We calculate the local field and the London free energy for this
geometry. Our calculation is a generalization of previous works
where the sample is taken as a semi-infinite superconductor.
As an application of this theory we investigate the flux spreading
at the superconducting surface.
\end{abstract}

\pacs{PACS numbers: 74.60.Ec, 74.60.Ge}


Scanning superconducting quantum interference device (SQUID)
microscope has been used to image interlayer Josephson vortices
trapped between the planes of layered superconductors. This
technique has been used to measure the out-plane London
penetration depth that gives the distance over which the
interlayer current $j_c$ changes as a function of in-plane
coordinates.\cite{moler98,kirtley98} These measurements have been
important to test the interlayer tunneling model as a candidate
to explain the mechanism of superconductivity for the high-$T_c$
superconductors.\cite{anderson98,chakravarty98}

Recently, Kirtley, Kogan, Clem, and Moler\cite{kirtley99} have
found expressions for the local magnetic field emerging from a
superconductor with the vortex lines parallel to the planes, and
normal to a crystal face. Their geometry consists of a
semi-infinite anisotropic superconductor. Furthermore, they have
used these expressions to fit the experimental data at the surface
in order to obtain an estimate of the value of the out-plane
penetration depth $\lambda_c$. They have shown that, neglecting
the vortex spreading at the surface may overestimate $\lambda_c$
as much as 30\%.

In the present paper we extend the work of Ref.~\onlinecite{kirtley99}
to an anisotropic layered superconducting film of finite thickness
and of infinite extent in the $bc$ face of the crystal. We will
show that, if the thickness of the film is of order or smaller
than $\lambda_c$, the magnetic field distribution
is even more affected by flux spreading.

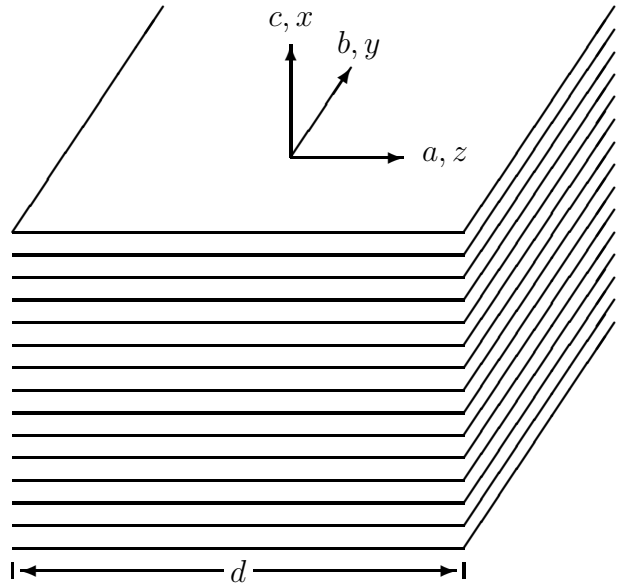
\begin{figure}
\caption{Geometry of the film used in this work. The vortex
lines are oriented perpendicular to the $bc$ face of the crystal.}
\label{fig1}
\end{figure}

Let us first formulate the problem to be solved. The geometry
we consider is illustrated in Fig.~\ref{fig1}.
We suppose that the vortex line is perpendicular to the film. We
will calculate the local field inside the film using the London
equation. For this geometry this equation is given by

\begin{equation}
\mbox{\boldmath $\nabla$}\times
[\stackrel{\leftrightarrow}{\mbox{\boldmath $\nabla$}}
\times{\bf h}]+{\bf h}=\hat{\bf z}\Phi_0\delta({\bf r})\;,
\label{eq_london}
\end{equation}
where $\stackrel{\leftrightarrow}{\mbox{\boldmath $\nabla$}}$
is the London
(tensor) penetration depth. This tensor is diagonal and its
components are given by
$\Lambda_{xx}=\Lambda_{yy}=\lambda_c^2$, $\Lambda_{zz}=
\lambda_{ab}^2$;
here $\lambda_{ab}$ and $\lambda_c$ are the in- and
out-plane penetration
depth respectively; $\Phi_0$ is the quantum flux. The film is
anisotropic along the $c$ direction.

Outside the sample, the local field satisfies the equation

\begin{equation}
\nabla^2{\bf h}=0\;.\label{eq_laplace}
\end{equation}

Although we will consider the case of a single vortex, the
generalization to the case of $N$ vortices is straightforward. To
proceed is more convenient to Fourier transform
Eqs.~(\ref{eq_london}) and (\ref{eq_laplace}). For $|z| < d/2$,
using the Maxwell equation $\mbox{\boldmath $\nabla$}\cdot{\bf
h}=0$, we obtain a set of three coupled differential equations
for the two dimensional Fourier transform of the local magnetic
field ${\bf h}({\bf k},z)=\int\,d^2r\,e^{-\i{\bf k}\cdot{\bf
r}}\,{\bf h}({\bf r},z)\;$,

\begin{equation}
\left [ 1+\lambda_{ab}^2k^2-\lambda_{ab}^2\frac{\partial^2}{\partial z^2}\right ]
h_x
 = 0\;,\label{eq_hx}
\end{equation}
\begin{equation}
\left [ 1+\lambda_{ab}^2k^2-\lambda_c^2\frac{\partial^2}
{\partial z^2}\right ] h_y
+(\lambda_c^2-\lambda_{ab}^2)ik_y\frac{\partial h_z}{\partial z}
 = 0 \;,\label{eq_hy}
\end{equation}
\begin{equation}
\left [ 1+\lambda_{ab}^2k_x^2+\lambda_c^2k^2_y-
\lambda_{ab}^2\frac{\partial^2}{\partial z^2}\right ] h_z
+(\lambda_c^2-\lambda_{ab}^2)ik_y\frac{\partial h_y}{\partial z}
=\Phi_0\;.\label{eq_hz}
\end{equation}

For $|z| > d/2$ one has

\begin{equation}
(\frac{\partial^2}{\partial z^2}-k^2){\bf h}=0\;.
\label{eq_laplace_fourier}
\end{equation}

At the vacuum-superconductor interfaces $z=\pm d/2$ the field
components are continuous and the component of the current
perpendicular to both film surfaces vanishes. One has

\begin{equation}
{\bf h}_<({\bf k},-d/2)={\bf h}_m({\bf k},-d/2)\;,\label{boundary<}
\end{equation}
\begin{equation}
{\bf h}_m({\bf k},d/2)={\bf h}_>({\bf k},d/2)\;,\label{boundary>}
\end{equation}
\begin{equation}
\hat{\bf z}\cdot\left [ {\bf D}_z({\bf k})
\times{\bf h}_m\right ] _{z=\pm d/2}=0\;,\label{boundary_current}
\end{equation}
\begin{equation}
{\bf D}_z({\bf k})\cdot{\bf h}=0\;,\label{eq_maxwell}
\end{equation}
where the operator
${\bf D}_z({\bf k})=i{\bf k}+\hat{\bf z}\frac{\partial}{\partial z}$. The
subscripts ($<,>$) stand for below the surface $z=-d/2$ and
above the surface $z=d/2$, respectively, whereas the subscript $m$
is meant for the field inside the sample.

We start by solving first Eq.~(\ref{eq_laplace_fourier}).
The solution which satisfies the boundary condition of
Eq.~(\ref{eq_maxwell}) takes the form

\begin{eqnarray}
{\bf h}_>({\bf k},z) & = & (-i{\bf k}+\hat{\bf z}k)
\varphi({\bf k})e^{-k(z-d/2)}\;,\label{h>} \\
{\bf h}_<({\bf k},z) & = & (i{\bf k}+\hat{\bf z}k)
\varphi({\bf k})e^{k(z+d/2)}\;,\label{h<}
\end{eqnarray}
where $\varphi({\bf k})$ is a scalar function which will
be determined by using the boundary condition either of
Eq.~(\ref{boundary<}) or (\ref{boundary>}).

Eq.~(\ref{eq_hx}) can also be easily solved. We have

\begin{equation}
h_{m,x}({\bf k},z)=W_1e^{\alpha z}+W_2e^{-\alpha z}\;,\label{sol_hx}
\end{equation}
where

\begin{equation}
\alpha=\sqrt{\frac{1+\lambda_{ab}^2k^2}{\lambda_{ab}^2}}\;,
\end{equation}
and the $W$'s are two constants to be determined by using
the boundary conditions.

The other two components of the local field can be determined
by decoupling Eqs.~(\ref{eq_hy}) and (\ref{eq_hz}). This can be done
by calculating the determinant of the matrix
formed by the coefficients of Eqs.~(\ref{eq_hy})
and Eq.~(\ref{eq_hz}). This yields the following equation
for $h_{m,y}$

\begin{equation}
\left [ 1+\lambda_{ab}^2k^2-\lambda_{ab}^2\frac{\partial^2}{\partial z^2}\right ]
\left [
1+\lambda_{ab}^2k_x^2+\lambda_c^2k_y^2-\lambda_c^2
\frac{\partial^2}{\partial z^2}
\right ] h_{m,y}=0\;.
\end{equation}

The solution for this equation is given by

\begin{equation}
h_{y,m}({\bf k},z)=W_3e^{\alpha z}+W_4e^{-\alpha z}
+W_5e^{\gamma z}+W_6e^{-\gamma z}\;,\label{sol_hy}
\end{equation}
where the $W$'s are constants to be determined
by using the boundary conditions and

\begin{equation}
\gamma=\sqrt{\frac{1+\lambda_{ab}^2k_x^2+\lambda_c^2k_y^2}
{\lambda_c^2}}\;.
\end{equation}

The solution for $h_{m,z}$ can be found by inserting Eq.~(\ref{sol_hy})
back into Eq.~(\ref{eq_hy}) or (\ref{eq_hz}).One has

\begin{equation}
h_{m,z}({\bf k},z)=\frac{\Phi_0}{\lambda_c^2\gamma^2}
+\frac{\alpha}{ik_y}(W_3e^{\alpha z}-W_4e^{-\alpha z})
-\frac{ik_y}{\gamma}(W_5e^{\gamma z}-W_6e^{-\gamma z})\;.
\label{sol_hz}
\end{equation}

The determination of the constants $W_i$ is very cumbersome and
we omit it here. We just present the main steps of the complete
solution. First of all, we use the Maxwell
equation (\ref{eq_maxwell}). This allows us
to write $W_3$ and $W_4$ in terms of $W_1$ and $W_2$.
Secondly, we use the boundary condition of
Eq.~ (\ref{boundary_current}) in both faces of the film. This leads
us to the solution of $W_5$ and $W_6$ in terms of
$W_1$ and $W_2$. Then, we are left only with three constants
to determine, namely, $W_1$, $W_2$, and $\varphi$. Thirdly, we use
the continuity of the local field at the film surfaces
[either Eq.~(\ref{boundary<}) or (\ref{boundary>}); both of them
yields the same solution to these constants]. One obtains,

\begin{eqnarray}
W_1 & = & -ik_x\frac{\varphi}
{2\sinh\left ( \frac{\alpha d}{2} \right )}
\;, \label{w1_final} \\
W_2 & = & ik_x\frac{\varphi}
{2\sinh\left ( \frac{\alpha d}{2} \right )}
\;, \label{w2_final} \\
W_3 & = & \frac{\lambda_{ab}^2k_xk_y}{1+\lambda_{ab}^2k_x^2}W_1\;,
\label{w3_final} \\
W_4 & = & \frac{\lambda_{ab}^2k_xk_y}{1+\lambda_{ab}^2k_x^2}W_2\;,
\label{w4_final} \\
W_5 & = & \frac{k_y}{k_x(1+\lambda_{ab}^2k_x^2)\sinh (\gamma d)}
\left \{ W_1\sinh\left [ \left (
\gamma+\alpha \right ) \frac{d}{2}\right ]
+W_2\sinh\left [ \left ( \gamma-\alpha \right )
\frac{d}{2}\right ]\right \} \;, \label{w5_final} \\
W_6 & = & \frac{k_y}{k_x(1+\lambda_{ab}^2k_x^2)\sinh (\gamma d)}
\left \{ W_1\sinh\left [ \left (
\gamma-\alpha \right ) \frac{d}{2}\right ]
+W_2\sinh\left [ \left ( \alpha+\gamma \right )
\frac{d}{2}\right ]\right \}\;, \label{w6_final} \\
\varphi({\bf k}) & = & \frac{\Phi_0}{\lambda_c^2\gamma^2}
\Delta({\bf k})\;, \label{phi_final}
\end{eqnarray}
where

\begin{equation}
\Delta({\bf k})=\left[ k+\frac{
\lambda_{ab}^2k_x^2\alpha\coth\left ( \frac{\alpha d}{2} \right ) +
\frac{k_y^2}{\gamma}\coth\left ( \frac{\gamma d}{2} \right )
}{1+\lambda_{ab}^2k_x^2}\right]^{-1}\;.\label{Delta}
\end{equation}

Finally, upon substituting Eqs.~(\ref{w1_final}-\ref{phi_final})
into Eqs.~(\ref{sol_hx}), (\ref{sol_hy}) and (\ref{sol_hz}), we
find for the local magnetic field inside the film

\begin{eqnarray}
h_{m,x}({\bf k},z) & = & -ik_x\varphi({\bf k})\frac{\sinh(\alpha z)}
{\sinh\left( \frac{\alpha d}{2}  \right)}\;,\label{hmx_final} \\
h_{m,y}({\bf k},z) & = & -ik_y\frac{\varphi({\bf k})}{1+\lambda_{ab}^2k_x^2}
\left [ \lambda_{ab}^2k_x^2\frac{\sinh(\alpha z)}
{\sinh\left( \frac{\alpha d}{2}  \right)}+
\frac{\sinh(\gamma z)}{\sinh\left( \frac{\gamma d}{2}  \right)}
\right ]\;,\label{hmy_final} \\
h_{m,z}({\bf k},z) & = & \frac{\Phi_0}{\lambda_c^2\gamma^2}
-\frac{\varphi({\bf k})}{1+\lambda_{ab}^2k_x^2}
\left [
\lambda_{ab}^2k_x^2\alpha\frac{\cosh(\alpha z)}
{\sinh\left( \frac{\alpha d}{2}  \right)}
+\frac{k_y^2}{\gamma}\frac{\cosh(\gamma z)}
{\sinh\left( \frac{\gamma
 d}{2}  \right)}
\right ]\;.\label{hmz_final}
\end{eqnarray}

We would like to point out that these results could not
be obtained from those of Ref.~\onlinecite{kirtley99} without
solving the problem. In fact, the solution of the
London equation for a superconducting film is different and
more difficult than for a semi-infinite superconductor.

Let us turn our discussion to the
calculation of the London free energy. The energy of the
vortex system is given by
$F=F_V+F_S$, where $F_V$ is the field energy in the vacuum
and $F_S$ is the energy inside the superconductor. One has

\begin{eqnarray}
F_V & = & \frac{1}{8\pi}\,\int\,\frac{d^2k}{(2\pi)^2}\,
\left \{
\int_{d/2}^{\infty}\,dz\,|h_>({\bf k},z)|^2
+\int_{-\infty}^{-d/2}\,dz\,|h_<({\bf k},z)|^2
\right \} \;,\label{freev} \\
F_S & = & \frac{1}{8\pi}\,\int\,\frac{d^2k}{(2\pi)^2}\,
\int_{-d/2}^{d/2}\,dz\,\left \{ |h_m({\bf k},z)|^2 \right .
\nonumber \\
& & +\left . \left [ {\bf D}_z({\bf k})\times{\bf h}_m({\bf k},z)
\right ] \cdot
\stackrel{\leftrightarrow}{\mbox{\boldmath $\nabla$}} \cdot
\left [ {\bf D}_z(-{\bf k})\times{\bf h}_m(-{\bf k},z)\right ]
\right \} \;.\label{frees}
\end{eqnarray}

By substituting the appropriate expressions of the local magnetic
field inside Eqs.~(\ref{freev}) and (\ref{frees}), after
a length algebra, we obtain

\begin{equation}
F=\frac{\Phi_0^2}{8\pi}\,\int\,\frac{d^2k}{(2\pi)^2}\,
\frac{1}{\lambda_c^2\gamma^2}\left [
d+2 \frac{\Delta({\bf k})}{\lambda_c^2\gamma^2}
\right ]\;.\label{free_final}
\end{equation}

The free energy can be generalized to an ensemble of
$N$ interacting vortex lines upon multiplying the
integrand of Eq.~(\ref{free_final}) by
$|S({\bf k})|^2$ where the structure factor is
given by

\begin{equation}
S({\bf k})=\sum_{i}\,e^{i{\bf k}\cdot{\bf R}_i}\;.
\end{equation}

Here ${\bf R}_i$ is the position of the $i$-vortex line.
Note that this extended result should be valid for an ensemble
of distorted vortices, that is, the positions of the
vortices do not necessarily correspond to the equilibrium
configuration.
The first term inside Eq.~(\ref{free_final}) represents
the interaction energy of the vortex lines as if
the surfaces were absent.
The second term represents the surface energy
associated to the magnetic energy of the stray field
at the superconductor-vacuum interface. Notice that for
$k$ small (large $r$), $\gamma^2\sim 1/\lambda_c^2$,
and $\Delta({\bf k})\sim 1/k$. Thus, the surface
energy goes as $\Phi_0^2/8\pi^2r$. Consequently, the
interaction on the surface depends neither on
the film thickness nor on the anisotropy. This is the Pearl result
for vortices emerging from a semi-infinite
isotropic supercondutor.\cite{pearl66}
Another interesting particular case of Eq.~(\ref{free_final}) is the
limit of a very thin film $d\rightarrow 0$, and $k$ small.
In this limit, from Eq.~(\ref{Delta}) it is straightforward
to show that
$\Delta({\bf k})=1/(k+2(\lambda_{ab}^2k_x^2+\lambda_c^2k_y^2)/d)$.
Therefore, from Eq.~(\ref{free_final}) we obtain

\begin{equation}
F=E_0\,\int\,\frac{dk^2}{(2\pi)^2}\,\frac{2\pi d}
{k\Lambda^{-1}+(k_x^2+\Gamma k_y^2)}\;,\label{energy_film}
\end{equation}
where $E_0=(\Phi_0/4\pi\lambda_{ab})^2$, $\Lambda=2\lambda_{ab}^2/d$,
and $\Gamma=\lambda_c^2/\lambda_{ab}^2$ is the anisotropy parameter.
This is precisely the energy of a single vortex in very thin film
first obtained by Pearl.\cite{pearl64,kogan93}

Now we will turn our attention to the streamlines
of the integrated field over $x$.
The distribution of magnetic field emerging on the surface
can be probed with a SQUID pickup loop. If the
SQUID probe is oriented in the $xy$ plane, the
total magnetic flux will be nearly equal to the pickup loop
size times\cite{kirtley99}

\begin{equation}
{\cal H}_z(y,z)=\int_{-\infty}^{\infty}\,h_z(x,y,z)\,dx=
\int_{-\infty}^{\infty}\,\frac{dk_y}{2\pi}\,h_z(0,k_y,z)e^{ik_yy}\;,
\label{hzcal}
\end{equation}
whereas, if the SQUID probe is oriented along the $xz$ plane,
the total magnetic flux is measured through the pickup loop
size times

\begin{equation}
{\cal H}_y(y,z)=\int_{-\infty}^{\infty}\,h_y(x,y,z)\,dx=
\int_{-\infty}^{\infty}\,\frac{dk_y}{2\pi}\,h_y(0,k_y,z)e^{ik_yy}\;.
\label{hycal}
\end{equation}

In order to compare our results with the results of
Ref.~\onlinecite{kirtley99}, we will replace the
vacuum-superconductor surfaces at $z=0$ and $z=-d$. This can be
done through the translation $z\rightarrow z+d/2$. From Eqs.~
(\ref{h>}), (\ref{h<}), and (\ref{hmx_final}-\ref{hmz_final})
we obtain,

\begin{eqnarray}
{\bf h}_>({\bf k},z) & = & (-i{\bf k}+\hat{\bf z}k)
\varphi({\bf k})e^{-kz}\;, \\
{\bf h}_<({\bf k},z) & = & (i{\bf k}+\hat{\bf z}k)
\varphi({\bf k})e^{k(z+d)}\;,
\end{eqnarray}

\begin{eqnarray}
h_{m,x}({\bf k},z) & = & -ik_x\varphi({\bf k})\frac{\sinh\left [
\alpha \left ( z+\frac{d}{2}\right ) \right ]}
{\sinh\left( \frac{\alpha d}{2}  \right)}\;, \\
h_{m,y}({\bf k},z)& =& -ik_y\frac{\varphi({\bf k})}{1+\lambda_{ab}^2k_x^2}
\left \{ \lambda_{ab}^2k_x^2\frac{\sinh\left [
\alpha \left ( z+\frac{d}{2}\right ) \right ]}
{\sinh\left( \frac{\alpha d}{2}  \right)}
+\frac{\sinh\left [
\gamma \left ( z+\frac{d}{2}\right ) \right ]}
{\sinh\left( \frac{\gamma d}{2}  \right)}
\right \}\;, \\
h_{m,z}({\bf k},z) & = & \frac{\Phi_0}{\lambda_c^2\gamma^2}
-\frac{\varphi({\bf k})}{1+\lambda_{ab}^2k_x^2}
\left \{
\lambda_{ab}^2k_x^2\alpha\frac{\cosh\left [
\alpha \left ( z+\frac{d}{2}\right ) \right ]}
{\sinh\left( \frac{\alpha d}{2}  \right)}
+\frac{k_y^2}{\gamma}\frac{\cosh\left [
\gamma \left ( z+\frac{d}{2}\right ) \right ]}
{\sinh\left( \frac{\gamma
 d}{2}  \right)}
\right \}\;.
\end{eqnarray}

The substitution of the appropriate expressions into
Eqs.~(\ref{hzcal}) and (\ref{hycal}) yields for the $z$
component of the $\vec{\cal H}$ field,

\begin{eqnarray}
{\cal H}^z_>(y,z) & = & \frac{\Phi_0}{\pi\lambda_c}\,
\int_0^{\infty}\,du\,\frac{\cos(y^{\prime}\sinh u)
\,e^{-z^{\prime}\sinh u}}
{\cosh u + \sinh u \coth \left ( \frac{d}{2\lambda_c}
\cosh u  \right )} \;, \\
{\cal H}_{m,z}(y,z) & = & \frac{\Phi_0}{\pi\lambda_c}\left \{
\frac{\pi}{2}e^{-|y^{\prime}|}-\int_0^{\infty}\,du\,
\tanh u \right . \nonumber \\
& & \left .
\times\frac{\cos(y^{\prime}\sinh u)}
{\cosh u + \sinh u \coth \left ( \frac{d}{2\lambda_c}
\cosh u  \right )}\frac{\cosh\left [
\left ( z^{\prime}+\frac{d}{2\lambda_c}  \right )\cosh u
\right ] }{\sinh \left ( \frac{d}{2\lambda_c}\cosh u \right )}
\right \}\;, \\
{\cal H}^z_<(y,z) & = & \frac{\Phi_0}{\pi\lambda_c}\,
\int_0^{\infty}\,du\,\frac{\cos(y^{\prime}\sinh u)
\,e^{(z^{\prime}+d/\lambda_c)\sinh u}}
{\cosh u + \sinh u \coth \left ( \frac{d}{2\lambda_c}
\cosh u  \right )} \;,
\end{eqnarray}
where $y^{\prime}=y/\lambda_c$ and $z^{\prime}=z/\lambda_c$.

The $y$ component takes the form

\begin{eqnarray}
{\cal H}^y_>(y,z) & = & \frac{\Phi_0}{\pi\lambda_c}\,
\int_0^{\infty}\,du\,\frac{\sin(y^{\prime}\sinh u)
\,e^{-z^{\prime}\sinh u}}
{\cosh u + \sinh u \coth \left ( \frac{d}{2\lambda_c}
\cosh u  \right )} \;, \\
{\cal H}_{m,y}(y,z) & = & \frac{\Phi_0}{\pi\lambda_c}\,
\int_0^{\infty}\,du\,
\frac{\sin(y^{\prime}\sinh u)}
{\cosh u + \sinh u \coth \left ( \frac{d}{2\lambda_c}
\cosh u  \right )}
\frac{\sinh\left [
\left ( z^{\prime}+\frac{d}{2\lambda_c}  \right )\cosh u
\right ] }{\sinh \left ( \frac{d}{2\lambda_c}\cosh u \right )}
\;, \\
{\cal H}^y_<(y,z) & = & -\frac{\Phi_0}{\pi\lambda_c}\,
\int_0^{\infty}\,du\,\frac{\sin(y^{\prime}\sinh u)
\,e^{(z^{\prime}+d/\lambda_c)\sinh u}}
{\cosh u + \sinh u \coth \left ( \frac{d}{2\lambda_c}
\cosh u  \right )} \;.
\end{eqnarray}

Note that in the limit of $d \rightarrow \infty$, our results
are exactly the same as those of Ref.~\onlinecite{kirtley99}.

The results for the $\vec{\cal H}$ field presented above should
be useful to interpret the experimental data obtained by using
scanning SQUID micorscopy. Unfortunately, the experiments
have been performed in samples of large thickness.
\cite{moler98,kirtley98} This renders
the test of the theory impracticable. In fact, vortices have been
magnetically imaged in films, but for a different geometry, that is,
the superconducing planes are taken parallel to the surfaces of
the film and the
vortex lines are considered perpendicular to the film surfaces.
\cite{kirtley99b}
In this case, we can extract the in-plane penetration depth
$\lambda_{ab}$ rather than $\lambda_c$, from the fitting
of the experimental data. So, we will restrict our
analysis only to the theoretical expressions.

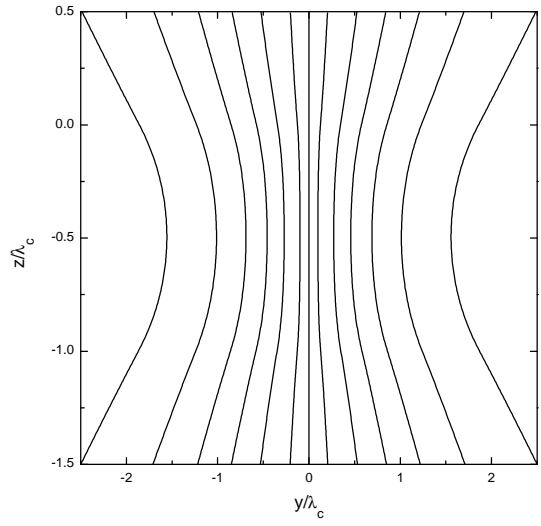
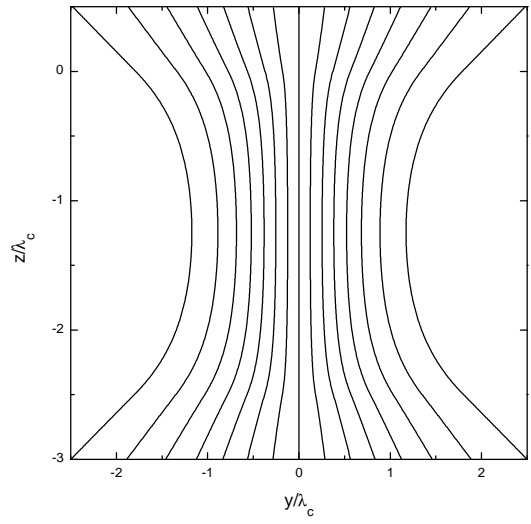
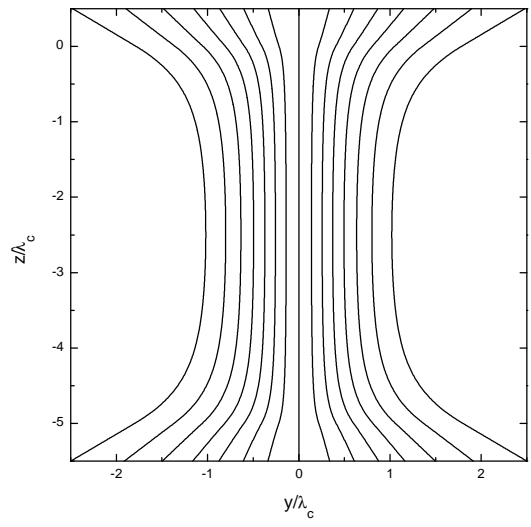
\begin{figure}
\caption{Streamline mapping of the integrated field
$\vec{\cal H}$, for an anisotropic superconducting film.
The spacing of the streamlines is proportional to
$(\partial {\cal H}_z/\partial y)^{-1}$ at $z=-(d+0.5\lambda_c)$.
\emph{(a)} shows the streamlines for a film of thickness
$d=\lambda_c$, \emph{(b)} for $d=2.5\lambda_c$, and
\emph{(c)} for $d=5\lambda_c$.}
\label{fig2}
\end{figure}

Fig.~\ref{fig2} shows the streamlines of the $\vec{\cal H}(y,z)$
field for a single interlayer vortex centered at $x=0$, $y=0$. The
streamlines were generated as sketched in
Ref.~\onlinecite{kirtley99}. We used various values of the
film thickness. Note that as the thickness of the film grows,
the flux spreading
is important only near the surface, whereas deep inside
the thinner film the streamlines are still very distorted,
except those close to the center of the vortex.

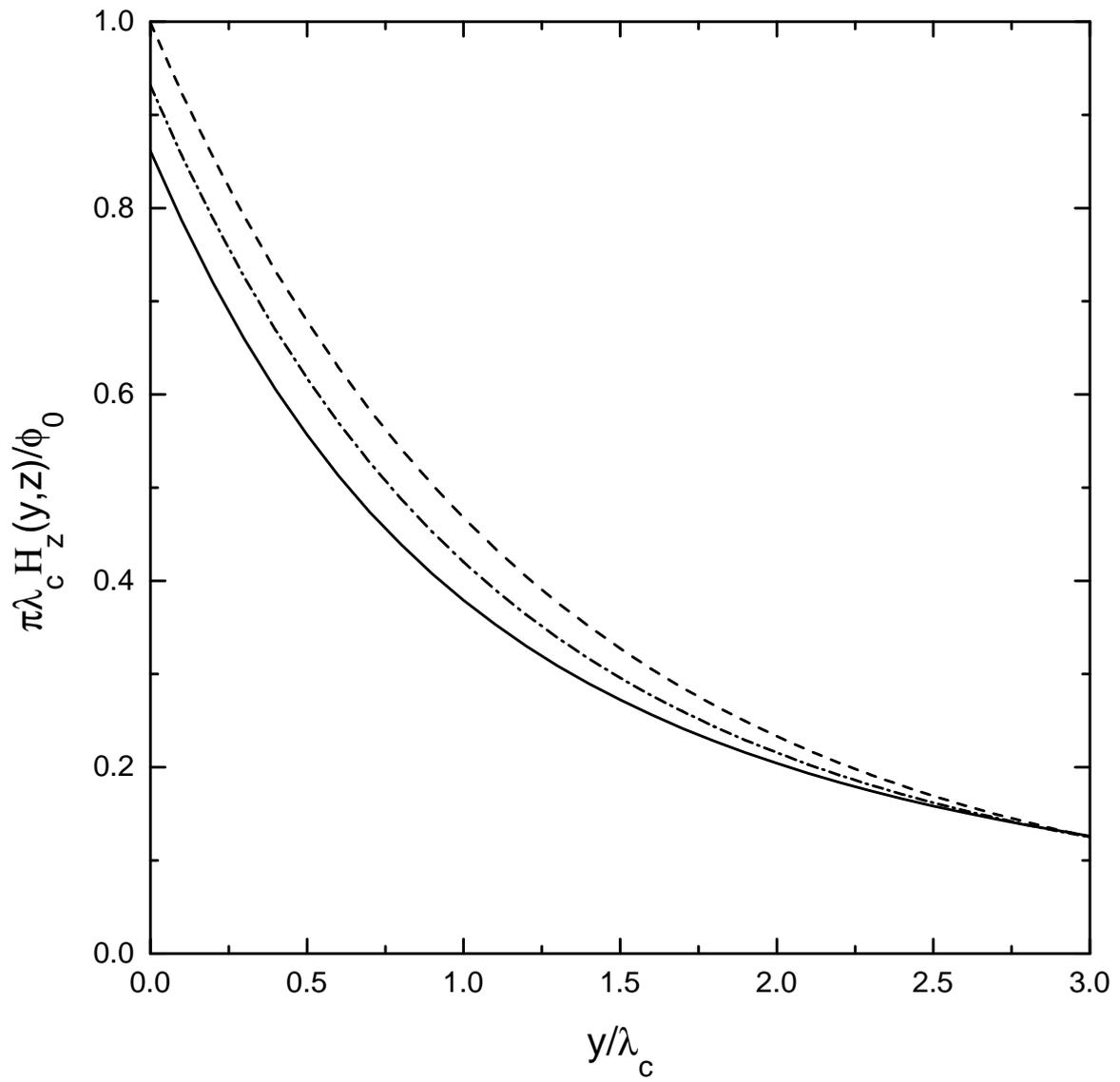
\begin{figure}
\caption{The $z$ component of the integrated field as a function of
$y$ at $z=0$ for three different values of the
film thickness: \emph{(a)}
$d=5\lambda_c$ (dashed line); \emph{(b)} $d=1.5\lambda_c$
(dot-dashed line); \emph{(c)} $d=\lambda_c$ (continuous line)
}\label{fig3}
\end{figure}

To see how important the flux spreading inside
a superconducting film is, we calculated numerically
$\pi \lambda_c{\cal H}_z(y,z)/\Phi_0$ as function of $y/\lambda_c$
for three different values of $d$ at $z=0$. As can be seen from
Fig.~\ref{fig3}, the full width at half maximum of the
flux contour is $1.87\lambda_c$ for the case $d=5\lambda_c$, while
it is $1.65\lambda_c$ for $d=\lambda_c$. Thus, if the flux spreading
inside the film is not taken into account, the value of
$\lambda_c$ could be underestimated by 10\%. This error grows
as the film thickness decreases.

Finally, we would like to point out that the present
results agree with their isotropic counterpart. If we set
$\lambda_{ab}=\lambda_c=\lambda$ in Eq.~(\ref{phi_final}) and
(\ref{Delta}), we obtain the same result as in Ref.~
\onlinecite{kirtley99b}.
Apparently, our results are different of those found in
Ref.~\onlinecite{gilson00}, but they show very similar streamlines.

In summary, we have calculated the field distribution of
a single vortex inside and
outside a layered superconducting film of arbitrary thickness.
We also calculated the London free energy of an ensemble of vortices.
From the expression for the energy one can recover
the interaction potential between vortices for a very
thin film\cite{pearl64,kogan93} and the vortices
emerging from a semi-infinite superconductor.\cite{kirtley99,pearl66}
In addition, we have shown that flux spreading inside
a superconducting film of order or smaller than $\lambda_c$
affects substantially the full width at half maximum of the
flux contour.

\acknowledgments
The author thanks the Brazilian Agencies FAPESP and CNPq
for financial support.

\newpage

\begin{center}
\setlength{\unitlength}{1cm}
\begin{picture}(8,6.5)(0,11)
\thicklines
\multiput(1.5,11)(0,-0.3){15}{\line(1,0){6}}
\multiput(7.5,11)(0,-0.3){15}{\line(2,3){2}}
\put(1.5,11){\line(2,3){2}}
\put(5.2,12){\vector(0,1){1.5}}\put(5.2,13.8){\makebox(0,0){$c,x$}}
\put(5.2,12){\vector(1,0){1.5}}\put(7.25,12){\makebox(0,0){$a,z$}}
\put(5.2,12){\vector(2,3){0.8}}\put(6.1,13.5){\makebox(0,0){$b,y$}}
\put(1.5,6.5){\line(0,1){0.1}}\put(1.5,6.5){\line(0,-1){0.1}}
\put(7.5,6.5){\line(0,1){0.1}}\put(7.5,6.5){\line(0,-1){0.1}}
\put(4.3,6.5){\vector(-1,0){2.7}}
\put(4.5,6.5){\makebox(0,0){$d$}}
\put(4.7,6.5){\vector(1,0){2.7}}
\end{picture}
\end{center}
\vspace*{8cm}
\begin{center}
{\Large Fig. 1/Sardella}
\end{center}

\begin{references}
\bibitem{moler98}K.\ A.\ Moler, J.\ K.\ Kirtley, D.\ G. Hinks,
T.\ W.\ Li, and M. Xu, Science {\bf 279}, 1193 (1998).
\bibitem{kirtley98}J.\ R.\ Kirtley, K.\ A.\ Moler, G.\ Villard,
and A.\ Maigman, \prl {\bf 81}, 2140 (1998).
\bibitem{anderson98}P.\ A.\ Anderson, Science {\bf 279},
1196 (1998).
\bibitem{chakravarty98}S.\ Chakravarty, Eur.\ Phys.\ J. B {\bf 5},
337 (1998).
\bibitem{kirtley99} J.\ R.\ Kirtley, V.\ G.\ Kogan, J.\ R.\ Clem
K. A. Moler, \prb {\bf 59}, 4343 (1999).
\bibitem{pearl66}J.\ Pearl, J.\ Appl.\ Phys.\ {\bf 37}, 4139 (1966).
\bibitem{pearl64}J.\ Pearl, Appl.\ Phys.\ Lett.\ {\bf 5}, 65 (1964).
\bibitem{kogan93}V.\ G.\ Kogan, A. Yu.\ Simonov, and M.\ Ledvij,
Phys.\ Rev.\ B {\bf 48}, 392 (1993). In this reference,
the anisotropic version of Pearl's vortex interaction
for very thin film has been found. Their result is identical
to Eq.~(\ref{energy_film}).
\bibitem{kirtley99b}J.\ R.\ Kirtley, C.\ C.\ Tsuei,
K. A.\ Moler, V.\ G.\ Kogan, J.\ R.\ Clem, and A. J. Turberfield,
Appl.\ Phys.\ Lett.\ {\bf 74}, 4011 (1999).
\bibitem{gilson00}G.\ Carneiro and E.\ H.\ Brandt, \prb {\bf 61},
6370 (2000); J.\ C.\ Wei and T.\ J.\ Yang, Jpn.\ J.\ Appl.\
Phys.\, Part 1 {\bf 35}, 5696 (1996).
\end{references}
\end{document}